\documentclass[prd,showpacs,amsmath,amssymb]{revtex4}

\usepackage{amssymb}
\usepackage{amsfonts}
\usepackage{graphicx}
\usepackage{keyval,graphicx}
\usepackage{textcomp,wasysym}
\newcommand{\be}{\begin{equation}}
\newcommand{\ee}{\end{equation}}
\newcommand{\bea}{\begin{eqnarray}}
\newcommand{\eea}{\end{eqnarray}}
\newcommand{\bean}{\begin{eqnarray*}}
\newcommand{\eean}{\end{eqnarray*}}

\begin{document}

\title{ Probing scalar meson structures in $\chi_{c1}$ decays into pseudoscalar and scalar}

\author{Qian Wang$^{1}$,  Gang Li$^{2}$, and Qiang Zhao$^{1,3}$}

\affiliation{1) Institute of High Energy Physics, Chinese Academy of
Sciences, Beijing 100049, P.R. China\\
2) Department of Physics, Qufu Normal University, Qufu, 273165, P.R.
China \\
3) Theoretical Physics Center for Science Facilities, CAS, Beijing
100049, China }

\vspace*{1.0cm}

\begin{abstract}

We evaluate the decay branching ratios of $\chi_{c1}\to PS$, in a
quark model parametrization scheme, where $P$ and $S$ stand for
pseudoscalar and scalar meson, respectively. An interesting feature
of this decay process is that the $c\bar{c}$ annihilate via the pQCD
hair-pin diagram is supposed to be dominant. Hence, this decay
process should be sensitive to the quark components of the
final-state light mesons, and would provide a great opportunity for
testing the mixing relations among the scalar mesons, i.e.
$f_0(1370)$, $f_0(1500)$ and $f_0(1710)$, by tagging the final state
pseudoscalar mesons.

\end{abstract}

\date{\today}





 \maketitle

\section{Introduction}

The usefulness of charmonium hadronic decays into light mesons is
that this transition occurs via a gluon-rich process. The initial
charm and anti-charm quark will annihilate into gluons and light
quarks will be produced in the final state through the hadronization
of intermediate gluons. For the interest of studying the structure
of the final state light mesons, especially in order to search for
signals for glueball candidates, the hadronic decays of charmonium
system provides an ideal platform on which the production of exotic
states can be correlated with a relatively well-understood state.
Such a tag sometimes exposes unexpected phenomena for which various
possible mechanisms can be examined. Review of heavy quarkonium
dynamics can be found in
Refs.~\cite{Brambilla:2004wf,Brambilla:2010cs}.

In the charmonium sector there are several recent observations
pertaining to scalar meson production in charmonium decays which
turn out to be unexpected. One is the BES-II results for $J/\psi\to
\phi f_0^i$~\cite{Ablikim:2004wn} and $\omega
f_0^i$~\cite{Ablikim:2004st}, ($i=1,2,3$ labels $f_0(1370)$,
$f_0(1500)$ and $f_0(1710)$), which show that the branching ratio of
$J/\psi\to \phi f_0(1710)$ is smaller than that of $J/\psi\to \omega
f_0(1710)$. In contrast, the branching ratio of $J/\psi\to \omega
f_0(1370)$ is smaller than $\phi f_0(1370)$. The paradox arising
here is that $f_0(1710)$ dominantly decays into $K\bar{K}$, hence is
usually believed to have a large $s\bar{s}$ component. Therefore,
one would naturally expect that the production of $f_0(1710)$
recoiled by the $\phi$ meson should be favored than recoiled by the
$\omega$ due to the OZI rule. Similar paradox occurs to the
$f_0(1370)$ which is strongly coupled to $4\pi$ and believed to be
dominated by a non-strange $q\bar{q}$ component. Efforts have been
made in the literature to explore the properties of these scalar
mesons and their
mixings~\cite{Amsler:1995tu,Amsler:1995td,Close:2000yk,Close:2005vf,Lee:1999kv,Cheng:2006hu}.

The other experimental observation comes from CLEO-c~\cite{:2008na}
and BES-II
measurements~\cite{bes-99b,bes-04,bes-05,bes-98i,bes-03c,bes-05b} of
$\chi_{c0,2}$ decays into meson pairs.  It shows that the decay
channels of $\chi_{c0,2}\to PP$ and $VV$ still respect the OZI rule
well and the DOZI processes are much
suppressed~\cite{Zhao:2007ze,Zhao:2005im}. As a consequence, the
leading transition amplitude is given by the singly OZI disconnected
(SOZI) transitions, and the decay branching ratios for
$\chi_{c0,2}\to PP$ and $VV$ still fit the pattern of SU(3) flavor
symmetry. In contrast, in the channel where scalar isoscalar $f_0$
states are produced, there are obvious deviations from the
expectation of the OZI rule. In particular, the BES-II
measurement~\cite{bes-05b} shows that one of the largest branching
ratios is from $\chi_{c0}\to f_0(1370) f_0(1710)$ for
$\chi_{c0,2}\to SS$ which greatly violate the OZI rule expectation.
As pointed out in Ref. \cite{Zhao:2007ze}, such a deviation could be
a strong evidence for the glueball-$q\bar{q}$ mixing in the scalar
isoscalar wavefunctions. Nevertheless, the scalar production is
correlated with the large OZI-rule breaking in the DOZI processes as
a common dynamic feature.

\begin{figure}
\begin{center}
\hspace{-1.5cm}
\includegraphics[scale=0.8]{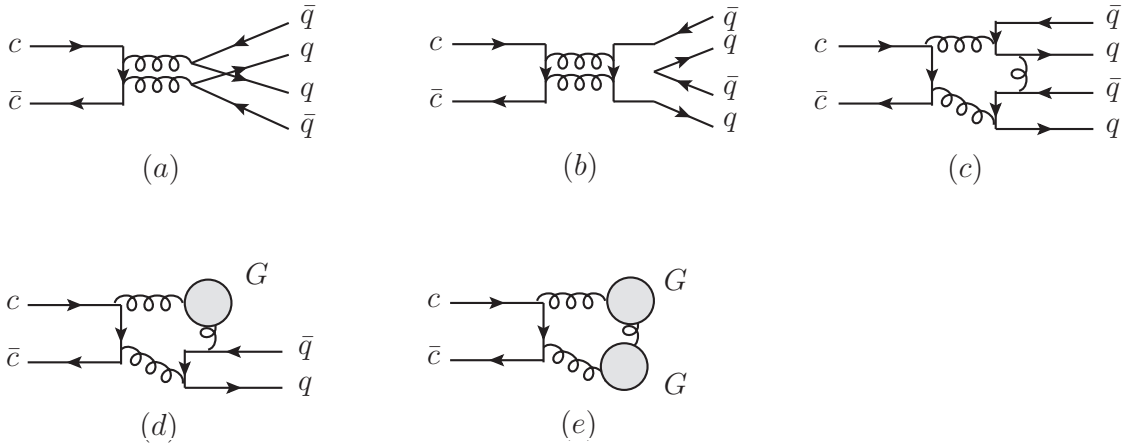}
\vspace{-15cm} \caption{The schematic Feynman diagrams of
$\chi_{c1}\to PS$ via two-gluon annihilations. }\label{fig-2g}
\end{center}
\end{figure}

Compared to $\chi_{c0,2}$ decays into two gluons, the two gluon
annihilations of the $\chi_{c1}$ in perturbative QCD is suppressed
by the Landau-Yang theorem~\cite{Yang:1950rg} in the on-shell limit
for those two gluons. As a result, the annihilations would be
dominated by the pQCD hair-pin diagram as shown in
Fig.~\ref{fig-2g}(a) instead of the connected diagram
(Fig.~\ref{fig-2g}(b)). As studied in Ref.~\cite{Huang:1996cs}, due
to the Landau-Yang theorem, the total width of $\chi_{c1}$ is
suppressed and nearly saturated by the hair-pin diagram where the
two gluons are not necessarily to be on shell simultaneously. The
strong suppression on the connected diagram
(Figs.~\ref{fig-2g}(b)-(e)) can be understood by the following
analyses. As a comparison, let us first take a look at the decays of
$\chi_{c0,2}\to 2g$, where Fig.~\ref{fig-2g}(b) generally plays an
important role~\cite{Zhou:2004mw}. In a view of quark-hadron
duality, this process can be treated as a two-step process, e.g.
$c\bar{c}(0^{++})\to 2g\to q\bar{q}(0^{++})$ and then the creation
of the second quark pair to form final-state hadrons. The first step
is analogue to a mixing process that the first light quark pair is
saturated by all the $q\bar{q}$ configurations with $J^{PC}=0^{++}$.
The second $q\bar{q}$ pair is produced via non-perturbative quark
pair creation for which the present calculations are generally in
the quark model framework~\cite{Zhou:2004mw}. The dominant
contributions from e.g. Fig.~\ref{fig-2g}(b) in $\chi_{c0,2}$ decays
are because of the enhancement of the first step transition, i.e.
$c\bar{c}(0^{++})\to 2g\to q\bar{q}(0^{++})$, when those two gluons
are both on shell. In contrast, the $\chi_{c1}$ decays via
Fig.~\ref{fig-2g}(b) with the on-shell gluons are forbidden by the
Landau-Yang theorem. Since the integrants of those gluon connected
diagrams drop quickly when the gluon momentum goes off shell, such a
critical constraint has strongly suppressed the contributions from
those gluon connected processes, i.e. Figs.~\ref{fig-2g}(b)-(e), as
a general character for the exclusive decays of $\chi_{c1}$. We note
that there is a gluon hair-pin diagram in association with
Fig.~\ref{fig-2g}(e) where the gluons must be off shell and can be
directly connected to the final state glueball components of the
pseudoscalar and scalar mesons. However, this process is relative
suppressed by the strong $\alpha_s$ in respect to
Fig.~\ref{fig-2g}(e) because of the additional three-gluon vertices.
Also, notice that the glueball components inside $\eta$ and $\eta'$
are rather small. We regard contributions from such a process as
subleading ones. In this work, we only consider $\eta$ and $\eta'$
as the flavor singlet and octet mixing states.

The following features can be further recognized: i)
Figure~\ref{fig-2g}(c) will suffer from OZI doubly disconnected
(DOZI) suppression  due to the large recoil momentum carried by the
exchanged gluon between the quarks in the final state. ii)
Figures~\ref{fig-2g}(d) and (e) may gain an enhancement by the gluon
powers considering that the gluon couplings to the glueball
generally do not pay a price. However, note that the total width of
$\chi_{c1}$ is nearly saturated by Fig.~\ref{fig-2g}(a), it is
confident to conclude that the DOZI and Landau-Yang suppression
still play a dominant role here. We caution that the neglect of
Figs.~\ref{fig-2g}(b)-(e) is based on qualitative argument and
experimental observations. Detailed model studies of those processes
are still needed to provide a quantitative prescription.

\begin{figure}
\begin{center}
\hspace{-1.5cm}
\includegraphics[scale=0.8]{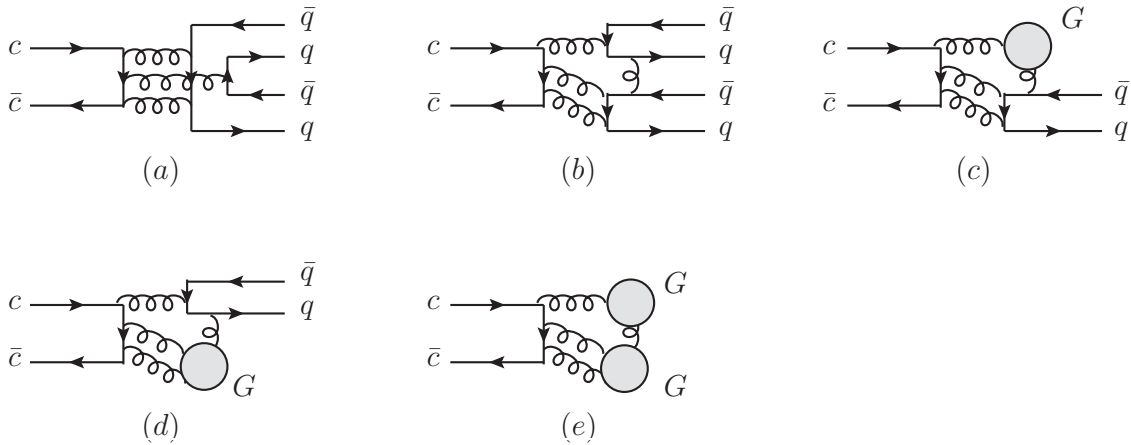}
\vspace{-15cm} \caption{The schematic Feynman diagrams of
$\chi_{c1}\to PS$ via three-gluon annihilations. }\label{fig-fey}
\end{center}
\end{figure}

The three gluon annihilations of $\chi_{c1}$ into $PS$ are also
highly suppressed. One reason is because of the increase of the
gluon powers. The other reason is that at least one of the final
state meson will be produced via higher-twist components in the
wavefunction as illustrated in Fig.~\ref{fig-fey}. Based on the
above considerations, it should be a reasonable approximation to
treat the hair-pin diagram as the leading contribution to
$\chi_{c1}\to PS$.

In this work, we will show that by tagging the quark components of
$\eta$ and $\eta'$ in the final state, it is possible to investigate
the quark components of the scalar mesons in $\chi_{c1}\to PS$. In
particular, this process turns out to be sensitive to the glueball
and $q\bar{q}$ mixing pattern. Therefore, it can be selective for
different mixing schemes and serve as an alternative way to study
the structure of scalar mesons.

As follows, we first give details of the parametrization scheme in
Sec. II. The numerical results are presented in Sec. III, and a
brief summary is given in the last section.

\section{Parametrization scheme }

A systematic parametrization scheme has been exploited for various
charmonium hadronic decays~\cite{Zhao:2005im,Zhao:2007ze,Li:2007ky},
where the SOZI and DOZI processes can be parameterized out based on
gluon counting rules. Those parameters can then be determined by
experimental data from which predictions can be made for unmeasured
channels. Early works based on similar parametrization can be found
in
Refs.~\cite{Amsler:1995tu,Amsler:1995td,Close:2000yk,Close:2005vf}.
In the case of $\chi_{c1}\to PS$, as discussed in the Introduction,
the dominant contribution is from Fig.~\ref{fig-2g}(a) while the
others are supposed to be strongly suppressed because of the
Landau-Yang theorem. It makes the parametrization rather simple as
we will detail below.

For the production of isoscalar pseudoscalar mesons, since it has
been well established that the glueball components inside $\eta$ and
$\eta'$ are rather small, it should be a good approximation to
neglect their possible internal glueball components, and simply take
the quark mixings in the SU(3) flavor basis:
\begin{eqnarray}
\eta&=&\cos\alpha_P|n\bar{n}\rangle-\sin\alpha_P|s\bar{s}\rangle,\\
\eta^\prime&=&\sin\alpha_P|n\bar{n}\rangle+\cos\alpha_P|s\bar{s}\rangle,
\end{eqnarray}
where $\alpha_P=\theta_P+\arctan\sqrt{2}$ and $\theta_P$ is usually
considered to be $-22^\circ\sim -13^\circ$. Here we adopt the
commonly used value $\theta_P=-19^\circ$ to evaluated the branching
ratios. With this mixing scheme, we eventually use the pseudoscalar
mesons $\eta$ and $\eta'$ as a flavor tag for the production of the
quark components of the scalar mesons via Fig.~\ref{fig-2g}(a),
while contributions from other processes can be neglected.

The rich spectrum of scalar mesons in the mass region of 1$\sim$ 2
GeV has initiated a lot of studies of the scalar mesons including
the search for the scalar glueball candidate (see
Refs.~\cite{Close:2002zu,Klempt:2007cp} and references therein). A
broadly discussed scenario is the glueball-$q\bar{q}$ mixing among
those three scalars, i.e. $f_{0}(1370)$, $f_{0}(1500)$ and
$f_{0}(1710)$. In the SU(3) flavor basis, these states as the
eigenstates of $n\bar{n}$, $s\bar{s}$ and $G$ components can be
generally expressed as
 \begin{eqnarray}
\left(
  \begin{array}{c}
    |f_{0}(1710)\rangle \\
    |f_{0}(1500)\rangle \\
    |f_{0}(1370)\rangle \\
  \end{array}
\right)= \hat{S}\left(\begin{array}{c}
               |G\rangle \\
               |s\overline{s}\rangle \\
               |n\overline{n}\rangle
             \end{array}
\right) =\left(\begin{array}{ccc}
                x_{1} & y_{1} & z_{1} \\
                x_{2} & y_{2}& z_{2}\\
               x_{3} & y_{3}& z_{3}
              \end{array}
\right)\left(\begin{array}{c}
               |G\rangle \\
               |s\overline{s}\rangle \\
               |n\overline{n}\rangle
             \end{array}
\right), \label{eq-scalar}
\end{eqnarray}
where $x_i$, $y_i$ and $z_i$ are the mixing matrix elements
determined by mixing mechanisms. Different models have different
solutions for the mixing
matrix~\cite{Close:2000yk,Close:2005vf,Amsler:1995tu,Amsler:1995td,Cheng:2006hu,Giacosa:2005zt}.
A critical difference among those mixing schemes focuses on the
magnitude of glueball components inside $f_0(1710)$ and $f_0(1500)$,
and in contrast, all those mixing schemes seem to agree that the
$f_0(1370)$ is dominated by the $n\bar{n}$ component. Unfortunately,
there still lack unique criteria for identifying the scalar glueball
state and distinguish those mixing schemes.

With $V_{a}$ standing for the potential of the SOZI process, a basic
transition parameter $g$ can be defined as
 \begin{eqnarray}
 g&\equiv &\langle(q\bar{q})_P(q\bar{q})_S\mid V_{a} \mid\chi_{c1}\rangle,
 \end{eqnarray}
where $q$ ($\bar{q}$) is a non-strange quark (antiquark).
Considering the SU(3) flavor symmetry breaking, which distinguishes
an $s$ quark pair production from the $u/d$ quarks in the
hadronizations, we introduce the SU(3) flavor symmetry breaking
parameter $R$,
\begin{eqnarray}
R&\equiv &\frac{\langle(s\bar{q})_P(q\bar{s})_S\mid V_{a}
\mid\chi_{c1}\rangle}{\langle(q\bar{q})_P(q\bar{q})_S\mid V_{a}
 \mid\chi_{c1}\rangle},
\end{eqnarray}
where $R=1$ is in the SU(3) flavor symmetry limit, while deviations
from unity implies the SU(3) flavor symmetry breaking. In general,
the value of parameter $R$ is around $R\simeq f_\pi/f_K=0.838$,
which provides a guidance for the SU(3) flavor symmetry breaking
effects.  For the creation of two pairs of $s\bar{s}$ via the SOZI
process, the recognition of the SU(3) flavor symmetry breaking in
the transition is
\begin{eqnarray}
R^{2}&=&\frac{\langle(s\bar{s})_P(s\bar{s})_S\mid V_{a}
\mid\chi_{c1}\rangle}{\langle(q\bar{q})_P(q\bar{q})_S\mid V_{a}
 \mid\chi_{c1}\rangle}.
\end{eqnarray}

Following the above parametrization rule, we can write down the
transition amplitudes for $I=0$ pseudoscalar ($\eta$ or $\eta'$) and
scalar meson ($f_0^i$, with $i=1,2,3$ for $f_0(1710)$, $f_0(1500)$
and $f_0(1370)$, respectively) pair production as the following
 \begin{eqnarray}
\langle \eta f_0^i|V_a |\chi_{c1} \rangle &=&g(z_i\cos\alpha_P-
y_i\sin\alpha_P R^2) {\cal F}({\bf P})
 \, \label{eta-f0}\\
\langle \eta' f_0^i|V_a |\chi_{c1} \rangle &=&g(z_i\sin\alpha_P+
y_i\cos\alpha_P R^2) {\cal F}({\bf P})\ .\label{etap-f0}
\end{eqnarray}
For $\chi_{c1}$ decays into other channels with $I\neq 0$, e.g. $\pi
a_0$ and $K \bar{K}_0^\ast+c.c.$, the transition amplitudes are
similar due to the exclusive contribution from Fig.~\ref{fig-2g}(a):
\begin{eqnarray}
 \langle \pi^{+} a_{0}^{-} |V_a|\chi_{c1}\rangle &=& \langle \pi^{-} a_{0}^{+}
 |V_a|\chi_{c1}\rangle =\langle \pi^{0} a_{0}^{0} |V_a|\chi_{c1}\rangle
 =g {\cal F}({\bf P})\ ,\label{a0pi0}\\
\langle K^{+} K^{*-}_{0} |V|\chi_{c1}\rangle &=&\langle K^{-}
K^{*+}_{0} |V|\chi_{c1}\rangle=\langle K^{0} \bar{K^{*}_{0}}^{0}
|V|\chi_{c1}\rangle=\langle \bar{K}^{0} K^{*0}_{0}
|V|\chi_{c1}\rangle=g R {\cal F}({\bf P}) \ .\label{k0k}
\end{eqnarray}
In the above equations, ${\cal F}({\bf P})$ is a commonly used form
factor defined as follows,
\begin{eqnarray}
 {\cal F}^2({\bf P}) \equiv |{\bf P}|^{2l} { \exp}(-{\bf
 P}^2/8\beta^2),
\end{eqnarray}
where ${\bf P}$ and $l$ are the three-vector momentum and the
relative orbit angular momentum of the final-state mesons,
respectively, in the $\chi_{c1}$ rest frame. We adopt $\beta=0.5$
GeV, which is commonly adopted in the
literature~\cite{Close:2005vf,Amsler:1995tu,Amsler:1995td,Close:2000yk}.
At leading order the decays of $\chi_{c1}\to PS$ are via $P$-wave,
i.e. $l=1$. This form factor accounts for the size effects arising
from the spatial wavefunction of the initial- and final-state mesons
in the hadronizations.

\section{Numerical results}

So far, the only available experimental information was given by
BES-II~\cite{Ablikim:2006vm,Nakamura:2010zzi}, i.e. $BR(\chi_{c1}\to
K_J^{\ast 0}(1430) \bar {K^0}+c.c.\to K_s^0 K^+\pi^- +c.c.) <8\times
10^{-4}$ and $BR(\chi_{c1}\to K_J^{*+}(1430)K^-+c.c.\to K_s^0
K^+\pi^- +c.c.) <2.3\times 10^{-3}$, where the statistics were
limited and the spin of $K_J^{*0}(1430)$ has not been determined. In
this process both $K_0^*(1430)$ and $K_2^*(1430)$ may have
contributions. However, notice that the leading hadronic
helicity-conserving amplitudes for $\chi_{c1}\to K_0^{\ast 0}(1430)
\bar {K^0}+c.c.$ and $K_2^{\ast 0}(1430) \bar {K^0}+c.c.$ are from
the $S_z=0$ components. It implies that the decay of $\chi_{c1}\to
K_2^{\ast 0}(1430) \bar {K^0}+c.c.$ via a $P$ wave will be
relatively suppressed by the Clebsch-Gordan coefficient, $\langle
20, 10|10\rangle^2=2/5$ in comparison with that of $K_0^{\ast
0}(1430) \bar {K^0}+c.c.$ This allows us to assume that the measured
branching ratio is dominated by $K_0^*(1430)\bar{K}+c.c.$, and set
up upper limits of branching ratios for other decay channels. In
another word, we can normalize the branching ratios of other decay
channels to the $K_0^{\ast 0}(1430) \bar {K^0}+c.c.$ channel (we
take the lower limit as a conservative estimate), and inspect the
variation of branching ratio fractions within a range of the SU(3)
flavor symmetry breaking parameter $R$. Notice that the
$K_0^*(1430)$ decay is nearly saturated by the $K\pi$ channel with
$BR(K_0^*(1430)\to K\pi)=(93\pm 10)\%$, which is larger than
$BR(K_2^*(1430)\to K\pi)=(49.9\pm 1.2)\%$~\cite{Nakamura:2010zzi}.
We can simply take into account the charge conjugate by a factor of
$3/2$ for $K_0^{\ast 0}(1430) \bar {K^0}+c.c.\to K_s^0 K^+\pi^-
+c.c.$ to obtain, $BR(\chi_{c1}\to K_J^{\ast 0}(1430) \bar
{K^0}+c.c.) < 1.2 \times 10^{-3}$ as the upper limit.

By studying this observable for different scalar meson mixing
schemes, we can identify criteria for the determination of the quark
contents of those scalar mesons. In turn, we expect to gain insights
into their structures.

From Eqs.~(\ref{eta-f0}) and (\ref{k0k}), we can take the branching
ratio fraction
\begin{equation}\label{ratio-eta}
\gamma_\eta\equiv \frac{BR(\chi_{c1}\to \eta f_0^i)}{BR(\chi_{c1}\to
K^0\bar{K}_0^{*0}+c.c.)}=\frac{|{\bf P}_\eta|(z_i\cos\alpha_P-
y_i\sin\alpha_P R^2)^2 {\cal F}^2({\bf P}_\eta)}{2|{\bf P}_K|R^2
{\cal F}^2({\bf P}_K)} \ ,
\end{equation}
where the factor 2 in the denominator is  due to the charge
conjugate factor for $K^0\bar{K}_0^{*0}+c.c.$ final states.
Similarly, for the $\eta'$ recoiling $f_0^i$ we can define
\begin{equation}\label{ratio-etap}
\gamma_{\eta'}\equiv \frac{BR(\chi_{c1}\to \eta'
f_0^i)}{BR(\chi_{c1}\to K^0\bar{K}_0^{*0}+c.c.)}=\frac{|{\bf
P}_{\eta'}|(z_i\sin\alpha_P+ y_i\cos\alpha_P R^2)^2 {\cal F}^2({\bf
P}_{\eta'})}{2|{\bf P}_K|R^2 {\cal F}^2({\bf P}_K)} \ ,
\end{equation}
and for the $\pi^0 a_0^0$ channel we have
\begin{equation}\label{ratio-pi}
\gamma_\pi\equiv \frac{BR(\chi_{c1}\to \pi^0 a_0^0)}{BR(\chi_{c1}\to
K^0\bar{K}_0^{*0}+c.c.)}=\frac{|{\bf P}_\pi|{\cal F}^2({\bf
P}_\pi)}{2|{\bf P}_K|R^2 {\cal F}^2({\bf P}_K)} \ .
\end{equation}

For $\chi_{c1}\to K^0\bar{K}_0^{*0}+c.c.$, the only parameter is $R$
for which $R\simeq f_\pi/f_K\sim 0.838$ is commonly adopted.
Applying the experimental upper limit, $BR(\chi_{c1}\to
{K_J^\ast(1430)}^0 \bar {K^0}+c.c.)<1.2\times
10^{-3}$~\cite{Ablikim:2006vm,Nakamura:2010zzi}, we can determine
the basic transition strength $g= 2.56\times 10^{-2}$ via
 \begin{eqnarray}\label{chi-kks}
 \Gamma(\chi_{c1}\to {K_J^\ast(1430)}^0 \bar {K^0}+c.c.) = \frac {|{\bf P}_K|g^2 R^2 {\cal F}^2({\bf
 P}_K)} {12\pi M_{\chi_{c1}}^2}.
 \end{eqnarray}
Also, the branching ratio of $\chi_{c1}\to \pi^0 a_0^0$ can be
estimated via Eq.~(\ref{ratio-pi}), i.e. $BR_{\chi_{c1}\to \pi^0
a_0^0}< 8.64\times 10^{-4}$.

Now we focus on Eqs.~(\ref{ratio-eta}) and (\ref{ratio-etap}) to
extract information about the scalar meson structures. The ratios
$\gamma_\eta$ and $\gamma_{\eta'}$ are now explicit functions of the
$q\bar{q}$ mixing elements. We will analyze three typical mixing
schemes in the literature. Predictions of the ratios would set up
criteria for future experimental tests of those mixing scenarios.

Scheme-I:

A systematic study by Close {\it et al} based on a perturbation
transition
mechanism~\cite{Amsler:1995tu,Amsler:1995td,Close:2000yk,Close:2005vf}
determines the glueball-$q\bar{q}$ mixing
matrix~\cite{Close:2005vf},
\begin{eqnarray}\label{Model-I}
 \hat{S}_1 &=&\left(\begin{array}{ccc}
                0.36 & 0.93 & 0.09 \\
                -0.84 & 0.35& -0.41\\
               0.40 & -0.07& -0.91
              \end{array}
\right).
\end{eqnarray}
In this scheme the $f_0(1710)$ is dominated by the $s\bar{s}$
component, but its glueball component is also sizeable. In contrast,
the $f_0(1500)$ is dominated by $G$ with a sizeable $s\bar{s}$
component. The $f_0(1370)$ is found dominated by the $n\bar{n} \
(\equiv (u\bar{u}+d\bar{d})/\sqrt{2})$.

\begin{figure}
\begin{center}
\hspace{-0.5cm}
\includegraphics[scale=0.65]{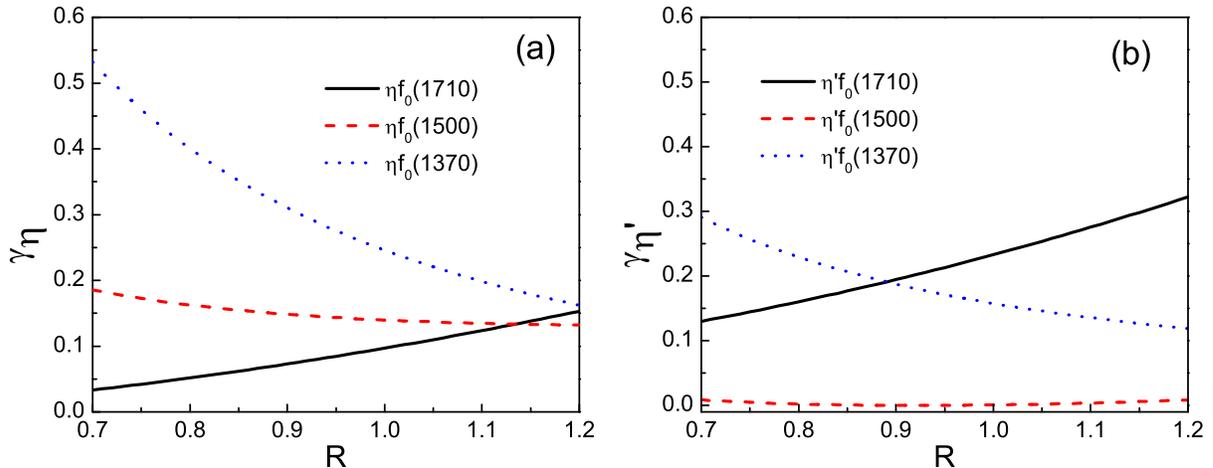}
\vspace{-7cm} \caption{The branching ratio fractions
$BR(\chi_{c1}\to \eta(\eta^\prime)f_0)/BR(\chi_{c1}\to
K^*_J(1430)^0\bar{K}^0+c.c.)$ as a function of the parameter R is
presented in this figure. In this case, the mixing matrix
$\hat{S}_1$ is used for scalar meson mixing. The solid line, dashed
line and dotted line are the branching ratio fractions of $\eta
f_0(1710)$, $\eta f_0(1500)$ and $\eta f_0(1370)$ in Diagram (a).
The cases in Diagram (b) are for $\eta^\prime f_0(1710)$,
$\eta^\prime f_0(1500)$ and $\eta^\prime f_0(1370)$.}\label{fig-s1}
\end{center}
\end{figure}

As mentioned earlier, the value $R=1$ corresponds to the SU(3)
flavor symmetry limit, while $R\simeq 0.838$ is the commonly adopted
SU(3) flavor symmetry breaking scale. We thus consider a small
variation of the parameter $R$ in the range of $R= 0.7\sim 1.2$, and
plot the $R$-dependence of branching ratio fractions $\gamma_\eta$
and $\gamma_{\eta'}$ in Fig.~\ref{fig-s1} with the mixing matrix
elements from Eq.~(\ref{Model-I}) for the mixing Scheme-I. It is
interesting to learn the following points about this scenario:

\begin{itemize}

\item Comparing the branching ratio fractions $\gamma_\eta$ and
$\gamma_{\eta'}$, we notice that the relative phases between the
mixing matrix elements critically determine the production strengths
of those scalars when recoil $\eta$ or $\eta'$.

\item Since the $n\bar{n}$ dominates the $f_0(1370)$ wavefunction and
the $s\bar{s}$ is negligibly small, the production rates of
$f_0(1370)$ are predicted to be larger than other channels in the
vicinity of $R\simeq 0.838$.

\item For the $f_0(1710)$, its production in association with $\eta$
is relatively suppressed by the significant cancelation between the
$s\bar{s}$ and $n\bar{n}$ components as shown by $\gamma_\eta\simeq
0.03\sim 0.15$. In contrast, its production with $\eta'$ is much
more enhanced with $\gamma_{\eta'}\simeq 0.13\sim 0.31$.

\item The $s\bar{s}$ and $n\bar{n}$ components are compatible in the
$f_0(1500)$, but out of phase. As a consequence, the production of
the $f_0(1500)$ seems to be unfavored in $\chi_{c1}\to PS$ for the
mixing Scheme-I. One also notices that both $\gamma_\eta$ and
$\gamma_{\eta'}$ are insensitive to $R$ as indicated by the dashed
curves.

\end{itemize}

\begin{figure}
\begin{center}
\hspace{-0.5cm}
\includegraphics[scale=0.65]{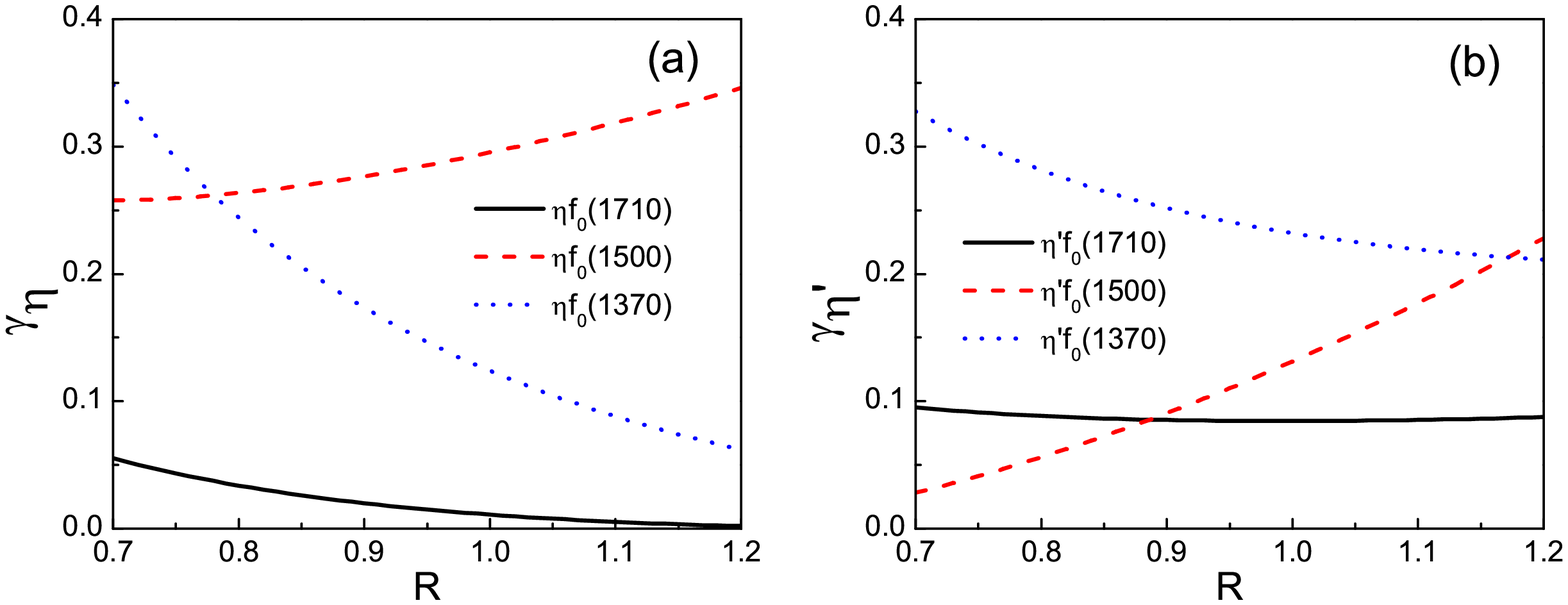}
\vspace{-7cm} \caption{The notations are similar to Fig.\ref{fig-s1}
with the mixing matrix $\hat{S}_2$. }\label{fig-s2}
\end{center}
\end{figure}

\begin{figure}
\begin{center}
\hspace{-0.5cm}
\includegraphics[scale=0.65]{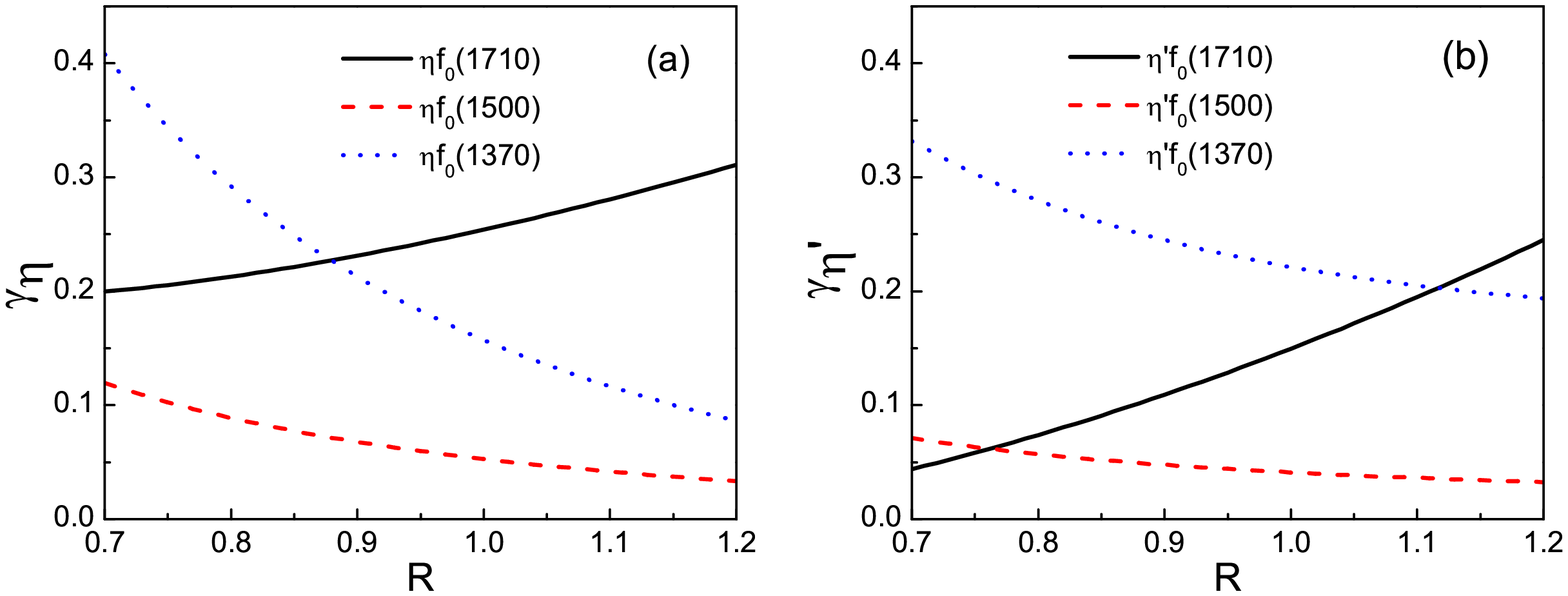}
\vspace{-7cm} \caption{The notations are similar to Fig.\ref{fig-s1}
with the mixing matrix $\hat{S}_{3a}$. }\label{fig-s3}
\end{center}
\end{figure}

\begin{figure}
\begin{center}
\hspace{-0.5cm}
\includegraphics[scale=0.65]{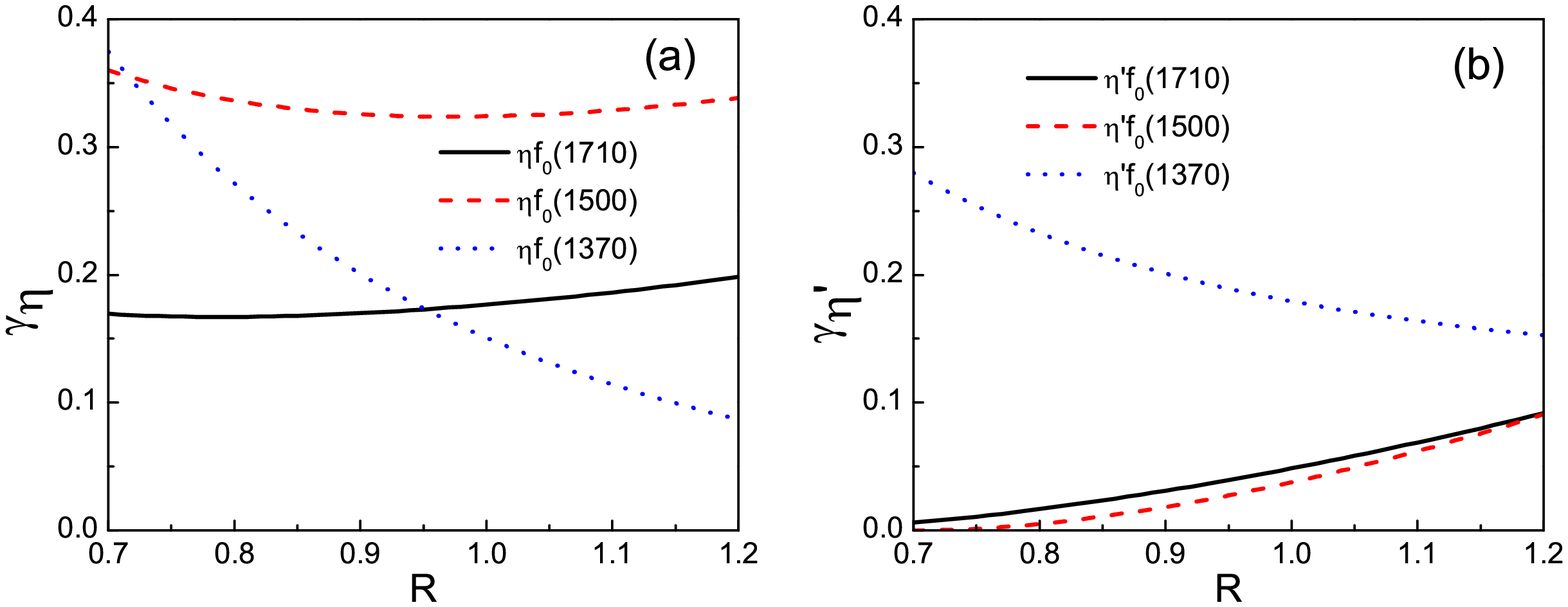}
\vspace{-7cm} \caption{The notations are similar to Fig.\ref{fig-s1}
with the mixing matrix $\hat{S}_{3b}$. }\label{fig-s4}
\end{center}
\end{figure}

Scheme-II:

The second mixing scheme was provided by Cheng, Chua and
Liu~\cite{Cheng:2006hu} based on the lattice QCD (LQCD) quenched
calculations by Lee and Weingarten~\cite{Lee:1999kv}. It was found
that $f_0(1710)$ is dominated by the  glueball component, while
$f_0(1500)$ and $f_0(1370)$ are dominated by $s\bar s$ and $n\bar
n$, respectively. This scenario is different from
Ref.~\cite{Zhao:2007ze} where the glueball-dominant state is the
$f_0(1500)$. The mixing scheme of Ref.~\cite{Cheng:2006hu} is
similar to that of Ref.~\cite{Lee:1999kv}. Therefore, we do not show
the numerical survey of these two solutions, but only present the
results for the typical solution from Ref.~\cite{Lee:1999kv}
\begin{eqnarray}\label{Model-CCL}
\hat{S}_{2}&=&\left(\begin{array}{ccc}
                0.859 & 0.302 & 0.413 \\
                -0.128 & 0.908& -0.399\\
               -0.495 & 0.290& 0.819
              \end{array}
\right) \ .
 \end{eqnarray}
The predicted branching ratio fractions $\gamma_\eta$ and
$\gamma_{\eta'}$ in terms of $R$ are plotted in Fig.~\ref{fig-s2}.
In comparison with Fig.~\ref{fig-s1}, the predicted decay pattern is
quite different in the range of $R>0.8$. In particular, one notices
the strong suppression of $\eta f_0(1710)$ in comparison with $\eta
f_0(1500)$ and $\eta f_0(1370)$ in the vicinity of $R\simeq 0.838$.
In the $\eta'$ production channels, the $\eta' f_0(1710)$ and $\eta'
f_0(1500)$ decays are relatively suppressed in comparison with
$\eta' f_0(1370)$.

Interestingly, one notices that in case of $R\simeq 0.7$, the
hierarchy of the branching ratio fractions between
Figs.~\ref{fig-s1} and \ref{fig-s2} are rather similar to each
other. In such a situation, one may need additional observables to
distinguish mixing schemes I and II.

Scheme-III:

Giacosa {\it et al.} obtained four possible solutions by fitting the
masses  and decay widths of those three $f_0$ states in an effective
chiral approach~\cite{Giacosa:2005zt}. Their two typical solutions,
i.e. $\hat{S}_{3a}$ and $\hat{S}_{3b}$ were obtained without direct
glueball decays, and have the following expressions,
 \begin{eqnarray}\label{solution-I}
\hat{S}_{3a}&=&\left(\begin{array}{ccc}
                -0.06 & 0.97 & -0.24 \\
                0.89 & -0.06& -0.45\\
               0.45 & 0.24& 0.86
              \end{array}
\right),
\end{eqnarray}
and
\begin{eqnarray}\label{solution-II}
\hat{S}_{3b}&=&\left(\begin{array}{ccc}
                -0.68 & 0.67 & -0.30 \\
                0.49 & 0.72& -0.49\\
              0.54 & 0.19& 0.81
              \end{array}
\right)\ .
 \end{eqnarray}
These two solutions were extracted with the OZI-rule violation
parameters $r=1.93\pm 0.29$ and $-2.07\pm 0.79$ determined in
$J/\psi\to \phi f_0^i$ and $\omega f_0^i$, and they are both of
order of one. Such a large OZI-rule violation parameter in
$J/\psi\to \phi f_0^i$ and $\omega f_0^i$ confirms the result of
Ref.~\cite{Close:2005vf}.  In $\chi_{c1}$ decays, as discussed in
the Introduction, the OZI rule and Landau-Yang suppression together
would lead to a small value for the OZI-rule violation parameter. It
means that we have neglected contributions from
Fig.~\ref{fig-2g}(d).

The main difference between these two solutions, $\hat{S}_{3a}$ and
$\hat{S}_{3b}$, lies in the different prescriptions of the glueball
and $s\bar{s}$ contents for the $f_0(1710)$ and $f_0(1500)$,
respectively. Compared with mixing scheme-I and II, it shows that
$\hat{S}_{3a}$ is similar to $\hat{S}_1$ of
Ref.~\cite{Close:2005vf}, but the second solution $\hat{S}_{3b}$ is
quite different.

In Figs.~\ref{fig-s3} and \ref{fig-s4}, the predicted branching
ratio fractions are presented for those two solutions $\hat{S}_{3a}$
and $\hat{S}_{3b}$, respectively. Their different prescriptions lead
to drastic changes of the production rate for the $f_0(1500)$ around
$R=0.838$. In the mixing scheme of $\hat{S}_{3a}$, the decay channel
$\eta f_0(1500)$ is suppressed in comparison with $\eta f_0(1710)$
and $\eta f_0(1370)$, while in $\hat{S}_{3b}$ it is strongly
enhanced to be larger than the other two channels. The $R$
dependence of $\gamma_\eta$ and $\gamma_{\eta'}$ can also be
observed in these two mixing schemes.

Interestingly, in the vicinity of $R\simeq 0.838$ the decay patterns
illustrated by those mixing schemes (four different mixing matrices)
can still be distinguished. As mentioned earlier that the structure
of mixing matrices, $\hat{S}_{3a}$  and $\hat{S}_1$, are similar to
each other, we can see that their predictions for the production of
$f_0(1710)$ are quite different. For instance, it shows that the
production of $f_0(1710)$ in association with $\eta$ is more favored
than with $\eta'$ in $\hat{S}_{3a}$, and it is opposite in
$\hat{S}_1$.

In Table~\ref{tab-branchingratio}, we list the branching ratios of
$\chi_{c1}\to \eta(\eta^\prime) f_0$ with $R=0.838$ as a predictions
from those mixing schemes. Combining what illustrated in
Figs.~\ref{fig-s1}-\ref{fig-s4}, we learn the following points
concerning the production of the scalars:

\begin{itemize}
\item The $f_0(1370)$ is dominated by the $n\bar{n}$ component
in all those schemes. Nevertheless, all those mixing schemes find
that the relatively small $s\bar{s}$ component is in phase to the
dominant $n\bar{n}$. As a consequence, the predicted branching
ratios of both $\eta f_0(1370)$ and $\eta' f_0(1370)$ turn out to
have a stable behavior.

\item For the $f_0(1500)$, sensitivities of the branching ratio
fractions $\gamma_\eta$ and $\gamma_{\eta'}$ to its quark contents
can be seen. The predicted branching ratios of $\chi_{c1}\to \eta
f_0(1500)$ are at the order of $10^{-4}$ which could be accessible
at BES-III. One also notices that the $\eta' f_0(1500)$ channel is
relatively suppressed in all those mixing schemes since the
$n\bar{n}$ and $s\bar{s}$ have opposite signs in the mixing schemes
$\hat{S}_1$, $\hat{S}_2$ and $\hat{S}_{3b}$, while in $\hat{S}_{3a}$
the $n\bar{n}$ and $s\bar{s}$ both are small.

\item The branching ratios of the $\eta f_0(1710)$ and $\eta'
f_0(1710)$ are also sensitive to the mixing schemes. Combine
together the decay patterns of other scalars, it is possible to
distinguish those mixing schemes in experiment.

\end{itemize}

\begin{table}
\caption{The upper limits of the branching ratios $\chi_{c1}\to \eta
f_0$ is presented with the SU(3) breaking parameter $R\equiv
f_\pi/f_K\simeq 0.838$. The experimental value $BR(\chi_{c1}\to
K^*_J(1430)^0\bar{K}^0+c.c.)<1.2\times
10^{-3}$~\cite{Nakamura:2010zzi} is used to predict the upper
limits.}
\begin{tabular}{ccccc}
  \hline\hline
  $BR(\chi_{c1}\to PS)$ & $\hat{S}_1$ & $\hat{S}_2$ & $\hat{S}_{3a}$ & $\hat{S}_{3b}$ \\\hline
  $\eta f_0(1710)$ & $7.2\times 10^{-5}$ & $3.3\times 10^{-5}$ & $2.63\times 10^{-4}$ & $2.03\times 10^{-4}$ \\
  $\eta f_0(1500)$ & $1.88\times 10^{-4}$ & $3.23\times 10^{-4}$ & $9.69\times 10^{-5}$ & $3.98\times 10^{-4}$ \\
  $\eta f_0(1370)$ & $4.32 \times 10^{-4}$ & $2.55\times 10^{-4}$ & $3.08\times 10^{-4}$ & $2.90\times 10^{-4}$ \\
  $\eta^\prime f_0(1710)$ & $2.10\times 10^{-4}$ & $1.04\times 10^{-4}$ & $1.04\times 10^{-4}$ & $2.70\times 10^{-5}$ \\
  $\eta^\prime f_0(1500)$ & $4.89\times 10^{-7}$ & $8.27\times 10^{-5}$ & $6.38\times 10^{-5}$ & $1.09\times 10^{-5}$ \\
  $\eta^\prime f_0(1370)$ & $2.51\times 10^{-4}$ & $3.21\times 10^{-4}$ & $3.17\times 10^{-4}$ & $2.61\times 10^{-4}$ \\
  \hline\hline
\end{tabular}
\label{tab-branchingratio}
\end{table}

It should also be pointed that experimental analysis may become much
more complicated due to the background contributions to the final
states. For the $PS$ channel, the final-state particles involve
$\eta\pi\pi, \ \eta K\bar{K}, \eta'\pi\pi$, and $\eta' K\bar{K}$
etc. As shown by the recent measurement from
CLEO-c~\cite{Adams:2011sq} that the $\eta\pi\pi$ and $\eta'\pi\pi$
channels may not be suitable for the search for the scalar meson
signals due to large background contributions. In contrast, the
$\eta K\bar{K}$ and $\eta' K\bar{K}$ channels may be more sensitive
to the scalar meson productions. With the high statistics
measurement at BEPC-II/BES-III, we anticipate that progress can be
made in the study of the scalar meson spectrum.

One should also be cautioned that the dominance of the hair-pin
diagram is a crucial assumption in this study. A better respect of
this assumption may be achieved in the bottomonium sector, namely,
in $\chi_{b1}\to PS$. Unfortunately, there are no experimental data
available at this moment for this channel. The future LHCb
experiment may be able to provide additional information about the
nature of those scalar mesons.

\section{Summary}

To summarize, we show that the decay of $\chi_{c1}\to PS$ could be
an ideal channel for probing the quark contents of the those scalar
mesons in the mass region of 1$\sim$2 GeV, i.e. $f_0(1370)$,
$f_0(1500)$ and $f_0(1710)$. Because of the suppression from the
Landau-Yang theorem to those gluon loop diagrams, this decay channel
at leading order should be dominated by the pQCD hair-pin transition
process. It thus allows us to tag the quark contents of the final
state scalars by the quark components of the recoiled $\eta$ and
$\eta'$. A prediction for the upper limits of the branching ratios
of $\chi_{c1}\to\eta f_0^i$ and $\eta' f_0^i$ can be made with the
available experimental data and based on different scalar mixing
schemes in the literature. It can be expected that a precise
measurement of $\chi_{c1}\to PS$ will be able to distinguish those
model prescriptions for the glueball-$q\bar{q}$ mixing scenario and
provide further evidence for the scalar glueball candidates.

\section{acknowledgement}

Q.Z. is indebted to K.-T. Chao, F.E. Close, M. Shepherd, J.-X. Wang,
and B.-S. Zou for useful discussions. Authors thank X.-Y. Shen, and
Z.-T. Sun for useful comments on an early draft. This work is
supported, in part, by National Natural Science Foundation of China
(Grant Nos. 11035006 and 10947007), Chinese Academy of Sciences
(KJCX2-EW-N01), Ministry of Science and Technology of China
(2009CB825200), and the Natural Science Foundation of Shandong
Province (Grant No. ZR2010AM011).

\end{document}